\documentclass[
]{ceurart}
\PassOptionsToPackage{hyphens}{url}\usepackage{hyperref}
\usepackage{listings}

\begin{document}

\copyrightyear{2022}
\copyrightclause{Copyright for this paper by its authors. Use permitted under Creative Commons License Attribution 4.0 International (CC BY 4.0).}


\title{Watching the Watchers: A Comparative Fairness Audit of Cloud-based Content Moderation Services}


\author[1,2]{David Hartmann}[%
email=d.hartmann@tu-berlin.de]
\address[1]{Faculty of Electrical Engineering and Computer Science, TU Berlin}
\address[2]{Weizenbaum Institute for the Networked Society}

\author[3]{Amin Oueslati}[%
email=amin.m.oueslati@gmail.com
]
\address[3]{Hertie School Berlin}

\author[1]{Dimitri Staufer}[%
email=staufer@tu-berlin.de]


\begin{abstract}
Online platforms face the challenge of moderating an ever-increasing volume of content, including harmful hate speech. In the absence of clear legal definitions and a lack of transparency regarding the role of algorithms in shaping decisions on content moderation, there is a critical need for external accountability. Our study contributes to filling this gap by systematically evaluating four leading cloud-based content moderation services through a \textit{third-party audit}, highlighting issues such as biases against minorities and vulnerable groups that may arise through over-reliance on these services. Using a black-box audit approach and four benchmark data sets, we measure performance in explicit and implicit hate speech detection as well as counterfactual fairness through perturbation sensitivity analysis and present disparities in performance for certain target identity groups and data sets. Our analysis reveals that all services had difficulties detecting implicit hate speech, which relies on more subtle and codified messages. Moreover, our results point to the need to remove group-specific bias. It seems that biases towards some groups, such as \textit{Women}, have been mostly rectified, while biases towards other groups, such as \textit{LGBTQ+} and \textit{PoC} remain.
\end{abstract}


\maketitle

\section{Introduction}
The digital age has brought a significant increase in online content. Worryingly, this also includes pernicious, unwanted content, such as hate speech \cite{bakalis2016regulating}. Online platforms responded by adopting extensive content moderation regimes \cite{degregorio2020democratising}. Absent a legal taxonomy of what constitutes hate speech, private companies are given substantial autonomy in their moderation practices, effectively making them the judges of public speech \cite{seering2020reconsidering, einwiller2020}.

Typically, large online platforms rely on so-called community guidelines against which speech is assessed. The assessment is done by human moderators, who are assisted by algorithms \cite{gorwa_algorithmic_2020}. The largest technology firms, such as Google, Microsoft, and Amazon, additionally offer content moderation as a service via cloud-based API access. While most organisations do not report the extent to which algorithms shape content moderation, the sheer amount of online speech makes reliance on algorithmic moderation inevitable \cite{Schluger2022ProactiveMO}.

To date, there exists no systematic evaluation of cloud-based content moderation services. The absence of public scrutiny is alarming, as open-source content moderation algorithms have continuously displayed biases against minorities and vulnerable groups \cite{garg_handling_2022, sap_social_2020, fortuna_toxic_2020, wiegand_detection_2019, hartvigsen_toxigen_2022, sheng_woman_2019}. This paper's contribution is twofold. Firstly, it offers the first comprehensive fairness assessment of four major cloud-based content moderation algorithms. Not only are these algorithms likely in use at the companies themselves, but they are also deployed by a vast number of smaller organisations through the SaaS model. Secondly, our auditing strategy may inform future bias audits of (cloud-based) content moderation algorithms. Importantly, our proposed approach solely assumes limited black-box access \cite{casper2024blackbox} and offers guidance on reinforced sampling strategies to achieve maximal scrutiny with limited resources, noting the realities of unsolicited audits from civil society organisations and academia \cite{Birhane2024AIAT, institute_algorithmic_2023, raji_outsider_2022}.

\section{Data and Method}
We gained researcher access to the Google Moderate Text API, Amazon Comprehend, Microsoft Azure Content Moderation, and the Open AI Content Moderation API. These services generate a hate speech score per text sequence, often split across several sub-categories, as well as a binary flag.
Our study uses the MegaSpeech, Jigsaw, HateXplain, and ToxiGen datasets \cite{pendzel_generative_2023, jigsaw_jigsaw_2019, mathew_hatexplain_2021, hartvigsen_toxigen_2022}. The selected datasets capture various forms of hate speech, with ToxiGen containing implicit and adversarial hate speech constructed around indirect messages \cite{elsherief_latent_2021}, while MegaSpeech and ToxiGen use generative AI to diversify speech corpora \cite{pendzel_generative_2023, hartvigsen_toxigen_2022}. Jigsaw and HateXplain contain human-written examples labeled by annotators, with MegaSpeech containing more hate speech corpora but no target group labels. MegaSpeech, HateXplain, and ToxiGen provide shorter text sequences, with on average 17.7, 23.3, and 18.1 words respectively, while Jigsaw is made up by longer sequences, 48.3 words on average.

We evaluate all cloud-based moderation algorithms across all datasets on a set of threshold-variant and threshold-invariant performance metrics \cite{elsafoury_bias_2023, borkan} at an aggregate level and also specifically for vulnerable groups.  
We ensure consistency across datasets by mapping these onto seven vulnerable groups (\textit{Women}, \textit{LGBTQ+}, \textit{PoC}, \textit{Muslim}, \textit{Asian}, \textit{Jewish}, \textit{Latinx}). Since MegaSpeech comes without labels, we train a Bi-LSTM model with the collected data set by \citet{yoder2022hate} (preliminary evaluation accuracy 78 \%) for target identity classification.
\begin{table}[t]
    \centering
    \scalebox{0.5}{
\begin{tabular}{l|ccccc||l|ccccc}
    \toprule
    Dataset & Moderation Service & ROC AUC & F1 & FPR & FNR &  Dataset & Moderation Service & ROC AUC & F1 & FPR & FNR \\
    \midrule
    \multirow{4}{*}{ToxiGen} & Amazon & \cellcolor{blue!20}70.4\% & \cellcolor{blue!20}68.9\% & \cellcolor{blue!20}7.2\% & 52.0\% & \multirow{4}{*}{MegaSpeech} & Amazon & 72.8 \% & 72.0 \% & 10.4 \% & \cellcolor{red!20}43.9 \%  \\
    & Google & 62.7\% & 62.7\% & \cellcolor{red!20}39.1\% & \cellcolor{blue!20}35.5\%  && Google & 73.3 \%   &  72.3 \% & \cellcolor{red!20}41.3 \% & \cellcolor{blue!25}12.0 \%\\
    & OpenAI & 70.3\% & 68.1\% & 33.2\% & 56.0\% &&OpenAI &\cellcolor{blue!25} 77.1 \% & \cellcolor{blue!25}76.7 \% & \cellcolor{blue!25}8.4 \% & 37.3 \%\\
    & Microsoft & \cellcolor{red!20}59.8\% & \cellcolor{red!20}57.4\% & 16.4\% & \cellcolor{red!20}64.0\% && Microsoft & \cellcolor{red!20}70.6 \% & \cellcolor{red!20}70.1 \% & 16.9 \% & 41.9 \% \\
    \hline
    \multirow{4}{*}{Jigsaw} & Amazon & \cellcolor{blue!20}92.2\% & \cellcolor{blue!20}92.2\% & \cellcolor{blue!20}7.5\% & 8.1\% & \multirow{4}{*}{HateXplain} & Amazon & 66.8\% & 66.25\% & 46.3 \% & \cellcolor{red!20}20 \%\\
    & Google & \cellcolor{red!20}69.9\% & \cellcolor{red!20}67.2\% & \cellcolor{red!20}58.4\% & \cellcolor{blue!20}1.8\% && Google & \cellcolor{red!20}52.2 \%   &  \cellcolor{red!20}58.9 \% & \cellcolor{red!20}78.2 \% & \cellcolor{blue!25} 4 \% \\
    & OpenAI & 78.6\% & 78.6\% & 17.1\% & 25.6\% && OpenAI &\cellcolor{blue!25} 72.9 \% & \cellcolor{blue!25}76.7 \% & \cellcolor{blue!25}45.4 \% & 8.86 \%  \\
    & Microsoft & 75.8\% & 75.7\% & 20.4\% & \cellcolor{red!20}28.1\%&& Microsoft & 63.1 \% & 60.2 \% & 63.6 \% & 10.3 \% \\
\end{tabular}
}
\caption{Performance metrics by moderation service and dataset. Blue shading signals the best performance, while red shading indicates the worst performance. ToxiGen includes 7,800 observations and HateXplain 14,000, while Jigsaw and MegaSpeech each contain 50,000. All datasets are balanced on toxic and non-toxic phrases.}
    \label{tab:peformance-metrics-aggreagte-outcomes}
\end{table}
At the group-level, we compute the pinned ROC AUC, a metric proposed by \citet{dixon_measuring_2018}, designed to provide a more robust measure for scale-invariant performance comparison across sub-groups. While this approach comes with its pitfalls, as the authors themselves note in a subsequent paper, it is the best scale-invariant metric to date when presented with group-level variation in biases \citep{borkan}.

Perturbation Sensitivity Analysis (PSA) offers an additional, arguably more robust evaluation of group-level biases by using counterfactual fairness evaluation\citep{prabhakaran-etal-2019-perturbation}. We follow prior research in defining an anchor group against which other groups are compared \citep{prabhakaran-etal-2019-perturbation}. Using the dominant majority group as baseline, Counterfactual Token Fairness (CFT) scores are computed as the difference in toxicity between the baseline and the corresponding minority group. PSA makes two assumptions: (1) counterfactual pairs should convey the same or neutral meaning, avoiding any implicit biases or derogatory connotations. While constructing toxic counterfactuals is theoretically possible, it is methodologically demanding and exceeds the scope of this project. Instead, we construct 34 \textit{neutral} counterfactual pairs. Importantly, each minority group is represented by multiple tokens, reflecting its different semantic representations. For instance, the minority group \textit{female} also manifests as \textit{woman} and \textit{women}. Furthermore, (2) there should be no unique interactions between a particular minority token and the context of the sentence that would skew the analysis. This is challenging in real-world applications, as certain combinations might evoke stereotypes or specific cultural connotations. Thus, the project uses data consisting largely of short and explicit statements. Furthermore, CFT scores are calculated separately for toxic and non-toxic statements, with the latter generally supporting the assumption of counterfactual symmetry more consistently.

PSA experiments are conducted using two distinct data sets. First, the synthetic \textit{Identity Phrase Templates} from \citet{dixon_measuring_2018} are used. The set contains 77,000 synthetic examples of which 50\% are toxic. These avoid stereotypes and complex sentence structures by design, which ensures that the symmetric counterfactual assumption is met. Mapping the dataset, which contains a broader set of identities, to the 34 minority token relevant to this study, results in 25,738 sentence pairs. Second, by applying the same logic, 9,190 sentence pairs are derived from the MegaSpeech dataset.

\section{Preliminary Results}
Table \ref{tab:peformance-metrics-aggreagte-outcomes} shows aggregated performance results for chosen benchmark data sets. Our results indicate notable disparities between moderation APIs. OpenAI's content moderation algorithm performs best for Megaspeech and Amazon Text Moderation on Jigsaw and ToxiGen, generalising well across data sets. On Jigsaw, Amazon Comprehend performs best. However, its near-optimal performance (92.2 \% ROC AUC) suggests that the Jigsaw data was likely included in training process of the Amazon Comprehend API . Overall, Google's API shows the worst performance across data sets. Its poor performance seems driven by a comparably high FPR, which suggests that the algorithm tends to overmoderate. In contrast,  Microsoft Azure Content Moderation is associated with a high FNR, suggesting it often misses hate speech.

Furthermore, all services struggle to detect implicit hate speech, reflected in their high False Positive Rates on ToxiGen. To this end, commercial moderation services do not fare much better than their open-source counterparts \citep{hartvigsen_toxigen_2022}. One likely cause is the limited availability of implicit hate speech datasets for training purposes. 

\begin{figure}[t]
\begin{minipage}[b]{0.45\linewidth}
    \centering
    \includegraphics[width=\linewidth]{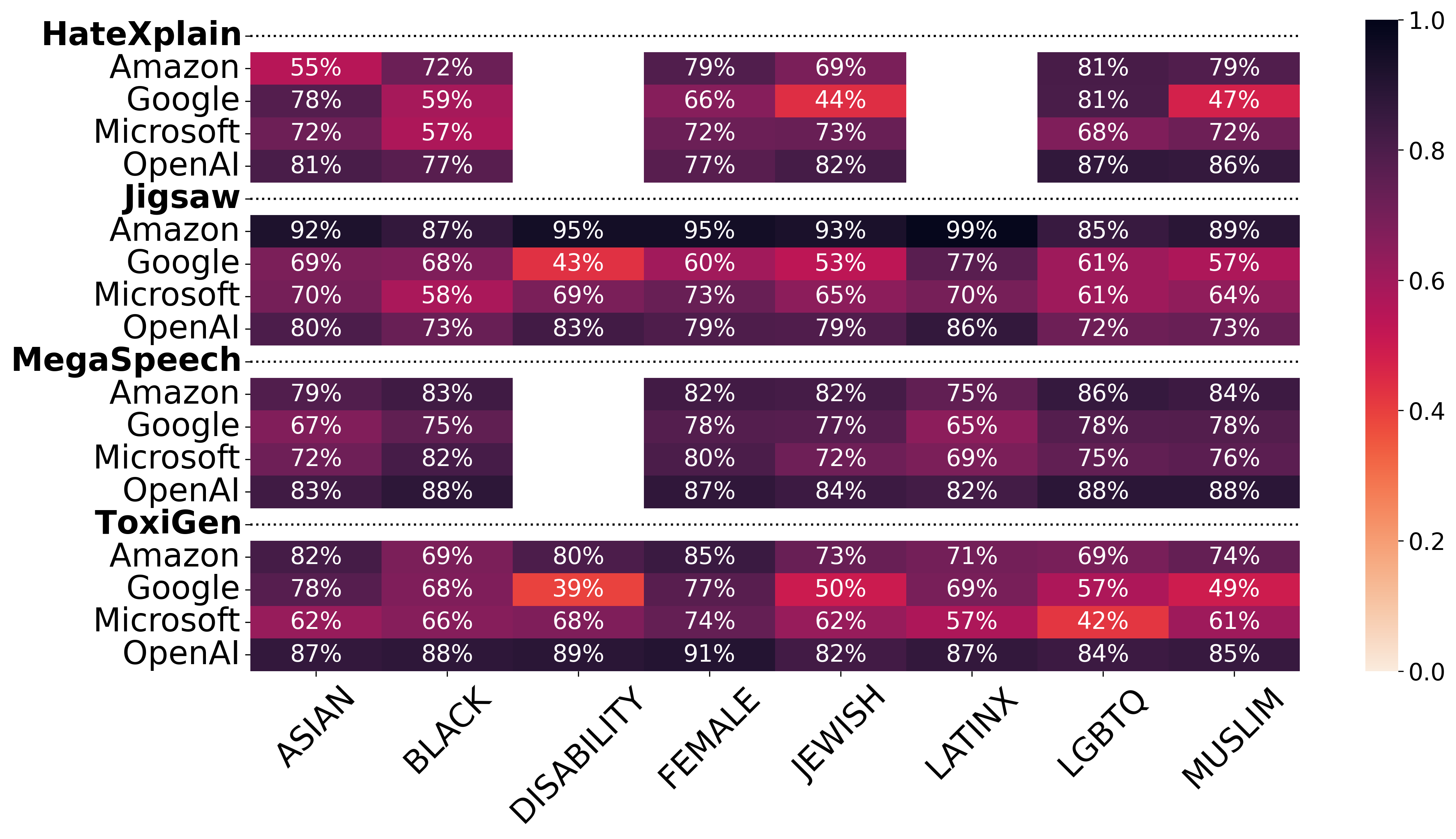}
    \label{fig:outcome1}
\end{minipage}
\hfill
\begin{minipage}[b]{0.45\linewidth}
   \includegraphics[width=\linewidth]{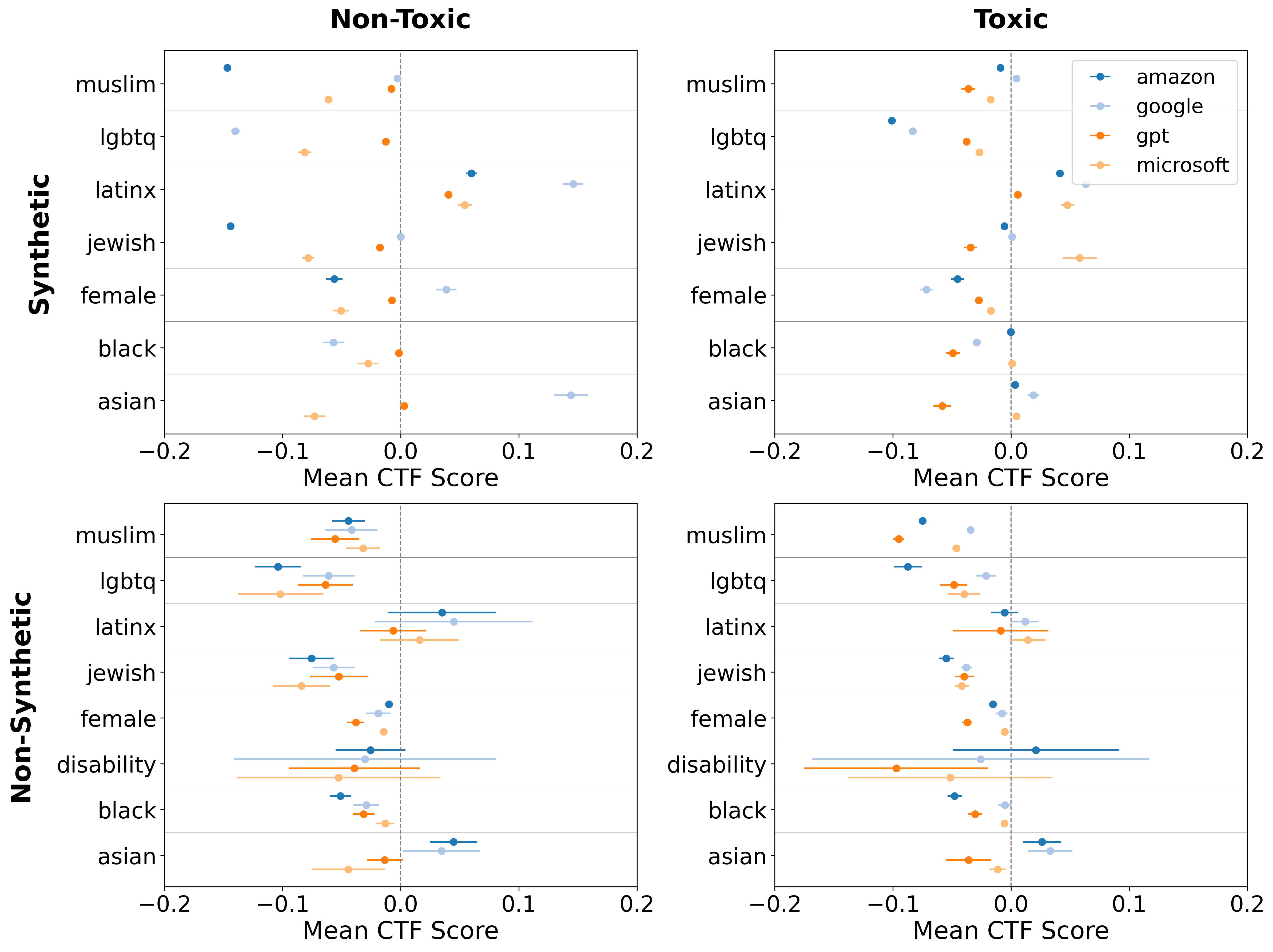}
\label{fig:outcome2}
\end{minipage}
\captionof{figure}{On the left, Pinned ROC AUC is presented by moderation service, dataset and minority group. ToxiGen includes 4,268 observations, HateXplain includes 1,748, Jigsaw consists of 19,228 observations and MegaSpeech is comprised of 33,886. On the right, CFT scores are visualized. They are computed through PSA on synthetic data from the Identity Phrase Templates in \citet{dixon_measuring_2018} and non-synthetic data from MegaSpeech, averaged per group and service, reported separately for non-toxic and toxic examples. Besides a point estimate, the figure also includes a 95\% confidence interval assuming a student-t distribution.}
\label{fig:performance}
\end{figure}

The comparative fairness evaluation of the identity group is presented via group-level pinned ROC AUC scores in Figure \ref{fig:performance}.\footnote{Due to space constraints, we only present one metric (ROC AUC). Future work includes a comprehensive analysis.} We find that all services tend to overmoderate speech concerning groups \textit{PoC} and \textit{LGBTQ+}. This is somewhat surprising as extensive prior research uncovered biases in open-source content moderation algorithms in relation to these groups \citep{garg_counterfactual_2019}. Commonly, such overmoderation occurs as toxic speech concerning these groups is overrepresented in the training data, and subsequently learned by the model. Most services fail to reliably detect hate speech aimed at groups \textit{Disability}, \textit{Asian}, and \textit{Latinx}. Lastly, the tendency of Google Text Moderation to overmoderate is puzzling but also alarming. While we cannot entirely rule out an error on our end, this observation is robust to different configurations of API sub-categories.

Figure \ref{fig:performance} (right) displays the PSA results. We find (1) differences in toxicity scores by and large are more pronounced on non-toxic than toxic data\footnote{ Intuitively this makes sense, as scores are generated non-linearly with a definite upper bound. Thus, when other elements in a sentence induce a high toxicity score, the marginal effect from identity tokens is comparably lower.} and (2) greater variation in the mean CFT scores in non-synthetic than in synthetic data\footnote{This was to be expected, as the sentences from MegaSpeech contain more contextual information that interacts with the tokens.}. 
Overall, the results suggest that most minorities are associated with higher levels of toxicity than dominant majorities, although these effects appear relatively small, and vary across groups and services. Group \textit{LGBTQ+} seems associated with the strongest negative bias, occurring for all samples and services. We observe limited negative bias against groups \textit{Latinx} and \textit{Asian}.

Summarizing, we uncovered both aggregate-level performance issues and group-level biases in major commercial cloud-based content moderation services. Importantly, while some shortcomings extend to all services, such as difficulty in detecting implicit hate speech or biases against group \textit{LGBTQ+}, others are confined to a particular service.

\bibliography{sample-ceur}

\appendix



\end{document}